\documentclass[11pt,twoside,a4paper,cmspaper,final,collab]{cms-tdr}
\def\svnVersion{2155c61-D}\def\svnDate{2026/07/20}\def\cmsCernNoTag{CERN-EP-2026-160}\def\cmsCernDate{\today}\def\cmsMessage{Submitted to Physics Letters B}
\begin{document}\cmsNoteHeader{HIN-24-011}

\ifthenelse{\boolean{cms@external}}
{
\newcommand{\cmsLeft}{upper\xspace}
\newcommand{\cmsRight}{lower\xspace}
\newcommand{\cmsLLeft}{Upper\xspace}
\newcommand{\cmsRRight}{Lower\xspace}
\newcommand{\cmsTable}[1]{#1}
}
{
\newcommand{\cmsLeft}{left\xspace}
\newcommand{\cmsRight}{right\xspace}
\newcommand{\cmsLLeft}{Left\xspace}
\newcommand{\cmsRRight}{Right\xspace}  
\newcommand{\cmsTable}[1]{\resizebox{\textwidth}{!}{#1}}
}

\newlength\cmsTabSkip\setlength{\cmsTabSkip}{1ex}

\newcommand{\gaga}{\PGg\PGg}
\newcommand{\tautau}{\ensuremath{\PGtp{}\PGtm{}}\xspace}
\newcommand{\tata}{\ensuremath{\PGt\PGt}}
\newcommand{\epem}{\ensuremath{\Pep{}\Pem{}}\xspace}
\newcommand{\mumu}{\ensuremath{\PGmp{}\PGmm{}}\xspace}
\newcommand{\ggee}{\ensuremath{\PGg\PGg \to \EE}\xspace}
\newcommand{\ggmumuFSR}{\ensuremath{\PGg\PGg \to \PGmp\PGmm\PGg}\xspace}
\newcommand{\ggmumu}{\ensuremath{\PGg\PGg \to \PGmp\PGmm}\xspace}
\newcommand{\ggtautau}{\ensuremath{\PGg\PGg \to \PGtp\PGtm}\xspace}
\newcommand{\PbggmmPbPure}{\ensuremath{\mathrm{Pb}\mathrm{Pb}\to\mathrm{Pb}(\PGg\PGg \to \PGmp\PGmm)\mathrm{Pb}}\xspace}
\newcommand{\muoneprong}{\ensuremath{\PGm+1\text{-prong}}\xspace}
\newcommand{\muthreeprong}{\ensuremath{\PGm+3\text{-prong}}\xspace}
\newcommand{\ethreeprong}{\ensuremath{\Pe+3\text{-prong}}\xspace}
\newcommand{\mue}{\ensuremath{\PGm+\Pe}\xspace}
\newcommand{\atau}{\ensuremath{a_{\PGt}}\xspace}
\newcommand{\mub}{\ensuremath{\mu\text{b}}\xspace}

\newcommand{\threeprong}{\ensuremath{\PGt_{3\text{-prong}}}\xspace}
\newcommand{\limitatau} {\ensuremath{-0.039<\atau<0.032}\xspace}
\newcommand{\resultxsection} {\ensuremath{\sigma_{\gaga\to\tata}^\text{fid}=555^{+60}_{-14}\mub}\xspace}
\newcommand{\resultxsectionSM} {\ensuremath{\sigma_{\gaga\to\tata}^\text{fid}=589^{+43}_{-40}\mub}\xspace}
\newcommand{\Aco}{\ensuremath{A_{\phi}}}
\newcommand{\gammaUPC}{{\texttt{gamma}-\textsc{upc}}}
\newcommand{\upcgen}{\textsc{Upcgen}}

\cmsNoteHeader{HIN-24-011}
\title{Measurement of the \texorpdfstring{\ggtautau}{gamma gamma to tau tau} cross section and constraints on the anomalous magnetic moment of the \texorpdfstring{\PGt}{tau} lepton in ultraperipheral PbPb collisions at \texorpdfstring{$\sqrtsNN = 5.02\TeV$}{sqrt(sNN) = 5.02 TeV}}

\date{\today}

\abstract{The production of $\PGt$ lepton pairs via photon-photon fusion, $\ggtautau$, is studied in ultraperipheral lead-lead collisions at a nucleon-nucleon center-of-mass energy of 5.02\TeV. The dataset, collected by the CMS experiment in 2018, corresponds to an integrated luminosity of 1.70\nbinv. Four different $\PGtp\PGtm$ decay final states are analyzed. A simultaneous likelihood fit to the measured lepton transverse momentum (\pt) distributions, which incorporates information from both spectral shape and normalization, is used to constrain the anomalous magnetic moment of the $\PGt$ lepton, $\atau$, and to extract the cross section of the process. The measured 95\% \CL interval for $\atau$ is $-0.039 < \atau < 0.032$. The fiducial cross section, $\resultxsection$ for tau leptons with $\pt^\PGt > 1\GeV$ and pseudorapidity $\abs{\eta^\PGt}<3$, is the most precise measurement for this process at the LHC to date and is in agreement with next-to-leading-order quantum electrodynamics predictions.}

\hypersetup{
pdfauthor={CMS Collaboration},
pdftitle={Measurement of the gamma-gamma to tautau cross section and constraints on the anomalous magnetic moment of the tau lepton in ultraperipheral PbPb collisions at 5.02 TeV},
pdfsubject={CMS},
pdfkeywords={CMS,  heavy ions, tau lepton}}

\maketitle

\section{Introduction}
\label{sec:Introduction}

Measurements of the magnetic moments of charged leptons ($g_{\Pell}$) provide stringent tests of quantum field theory and the standard model (SM) of particle physics. A small but nonzero anomalous magnetic moment, $a_{\Pell} \equiv (g_{\Pell}-2)/2$, arises because of quantum corrections to the tree-level value of $g_{\Pell}=2$, expected for a spin-$1/2$ particle. For the electron, the SM prediction agrees with experiment to better than one part per billion~\cite{Aoyama:2019ryr,Fan:2022eto}, while measurements of the muon $a_\PGm$ have shown tensions with respect to certain theoretical calculations~\cite{Muong-2:2023cdq,Aoyama:2020ynm,Aliberti:2025beg}. The tau lepton, with its larger mass, is expected to be more sensitive to potential new physics effects contributing to $\atau$ through virtual loop corrections~\cite{Crivellin:2021spu}. However, the short tau lepton lifetime precludes direct measurements of $\atau$ in storage rings. Instead, $\atau$ is best constrained via photon-fusion production of tau lepton pairs, $\ggtautau$, in high-energy collisions~\cite{delAguila:1991rm,Cornet:1995pw,Atag:2010ja,Beresford:2019gww,Dyndal:2020yen}. Early limits were set at the CERN LEP collider~\cite{DELPHI:2003nah}, followed by the CERN LHC measurements in ultraperipheral collisions (UPCs) of heavy ions~\cite{ATLAS:2022ryk,CMS:2022arf,ATLAS:2026wrz} and protons~\cite{CMS:2024qjo}. The most precise result today, $\atau = 0.0009^{+0.0032}_{-0.0031}$~\cite{CMS:2024qjo}, remains about a factor of three less precise than required to begin probing the SM prediction, $\atau^\text{SM}= 0.00117721(5)$, which is dominated by quantum electrodynamics (QED) contributions. The number in parentheses denotes the uncertainty of the last quoted digit~\cite{Eidelman:2007sb}.

\begin{figure}[htbp]
\centering
 \includegraphics[width=0.5\textwidth]{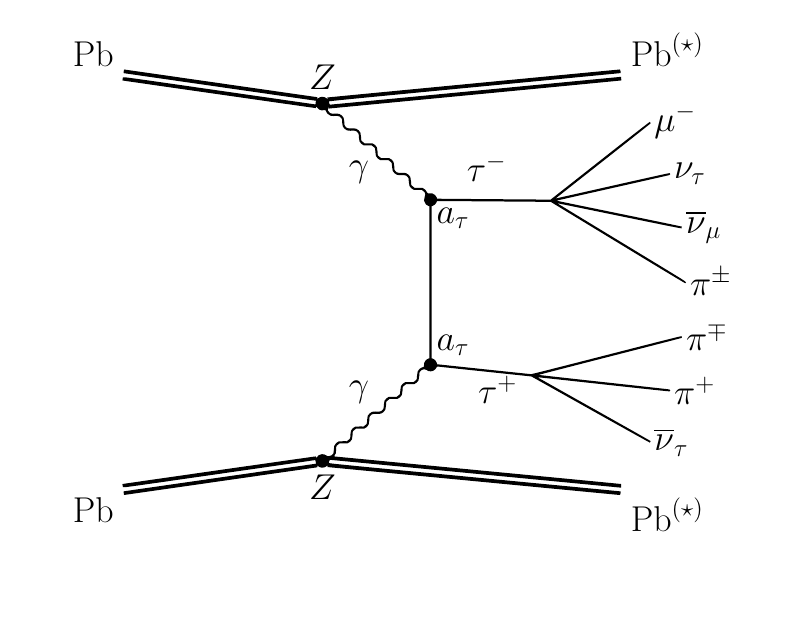}
\caption{Diagram of the $\text{PbPb}\xrightarrow{\gaga}\Pb^{(*)}\tautau\Pb^{(*)}$ process, with one tau lepton decaying leptonically and the other hadronically (3-prong final state). The charge-conjugate mode with a positive muon is implicitly included. The two $\PGg\PGt\PGt$ vertices are sensitive to $\atau$. The star superscript in $\Pb^{(*)}$ indicates a possible electromagnetic excitation of the lead ion.}
\label{fig:signal_mu3prong-Feynman}
\end{figure}

The measurement of $\atau$ via the $\ggtautau$ process in UPCs of heavy ions, where the impact parameter of the collision exceeds twice the nuclear radius~\cite{Baltz:2007kq}, offers several advantages over the same process measured in electron-positron ($\epem$) and proton-proton ($\Pp\Pp$) collisions. First, the cross section at low $\tautau$ invariant masses is enhanced by approximately a factor of $Z^4$, where $Z$ is the electric charge, which amounts to $Z=82$ for a lead (Pb) nucleus. Second, compared to $\Pp\Pp$ collisions, the absence of simultaneous interactions within the same bunch crossing (``pileup'') in the heavy ion operation mode at the LHC leads to very clean final states, containing only the leptonic and/or hadronic decay products of the $\PGt$ leptons (Fig.~\ref{fig:signal_mu3prong-Feynman}). Third, the photons emitted by Pb nuclei in UPCs are quasireal, whereas they have larger virtualities in $\Pp\Pp$ and $\epem$ collisions, and the two exclusively produced tau leptons are thus emitted nearly back-to-back. The production cross section and the kinematic distributions of the \PGt leptons and their decay products in \ggtautau events depend on the value of \atau. Measurements of these observables therefore facilitate placing experimental constraints on the anomalous magnetic moment of the $\PGt$ lepton. This extends the use of PbPb collisions beyond traditional heavy ion physics topics~\cite{CMS:2024krd}, enabling the study of QED processes under experimental conditions as clean as those in $\epem$ collisions.

The final-state signature of interest is the exclusive production of two tau leptons, $\text{PbPb}\xrightarrow{\gaga}\Pb^{(*)}\tautau\Pb^{(*)}$, where their decay products are measured in the central part of the detector and the outgoing Pb ions survive the interaction and escape undetected at very low angles with respect to the beam. The lead ions can also potentially exchange soft photons and be excited above their ground state, as denoted by the $(*)$ superscript. Such an excitation is often followed by the emission of one or a few neutrons collinear to the beam(s), a process called electromagnetic dissociation (EMD)~\cite{Baur:2003gv}.

The first CMS observation of the $\ggtautau$ process used PbPb collisions at a nucleon-nucleon center-of-mass energy of $\sqrtsNN = 5.02\TeV$, collected in 2015 and corresponding to an integrated luminosity of 0.4\nbinv~\cite{CMS:2022arf}. This first study considered only the \muthreeprong final state, in which one tau lepton decays into a charged muon and the other decays hadronically into three charged hadrons (3 prongs) as illustrated in Fig.~\ref{fig:signal_mu3prong-Feynman}. The present analysis extends this initial study by including three additional $\PGtp\PGtm$ final states (\muoneprong, \mue, and \ethreeprong), by increasing the integrated luminosity 
to $1.70\nbinv$, and by introducing a simultaneous fit that combines the normalizations and shapes of the decay-lepton kinematic distributions to extract a more precise fiducial cross section and tighter constraints on $\atau$.

The measurement is performed in a fiducial phase space defined by the tau leptons transverse momenta $\pt^\PGt > 1\GeV$ and pseudorapidities $\abs{\eta^\PGt} < 3$, with all four $\PGtp\PGtm$ final states contributing to the signal via different decay signatures. Different selection criteria are applied to suppress backgrounds arising from other UPC processes---such as photon--photon, photon--nucleus, and photon--pomeron interactions producing low-multiplicity final states---that can mimic the exclusive ditau signal of interest. The \muoneprong final state has the largest $\tautau$ branching fraction ($\mathcal{B} \approx 17.4\%$) and provides the highest expected sensitivity, despite being affected by larger background contamination as compared with other channels. The \muthreeprong channel ($\mathcal{B} \approx 5.1\%$) takes advantage of the excellent muon identification capabilities of CMS, with backgrounds mimicking 3-prong tau lepton decays being efficiently suppressed. Electron final states are generally less favorable than their muon counterparts because the higher electron \pt trigger thresholds and lower reconstruction efficiencies reduce the available signal yields. For this reason, the $\Pe+1\text{-prong}$ channel is not used, as its substantial background would significantly limit the achievable sensitivity. The \ethreeprong final state ($\mathcal{B}\approx 5.2\%$) remains sufficiently populated to be included in the study and is treated analogously to the \muthreeprong channel, though with a lower expected signal yield for the same reasons. The \mue channel ($\mathcal{B} \approx 6.2\%$) has the lowest background contamination among all channels, but also suffers from relatively lower yields because of the small branching fraction and large electron \pt trigger thresholds. The resulting limits on $\atau$ improve upon the previous CMS PbPb result~\cite{CMS:2022arf} by more than a factor of three, and the measured cross section---the most precise to date for this process at the LHC---is compared with next-to-leading-order (NLO) QED predictions~\cite{Shao:2024dmk,Shao:2025bma}. Tabulated results are provided in the HEPData record~\cite{HEPData}.

\section{The CMS experiment}
\label{sec:CMS}

The CMS apparatus~\cite{CMS:2008xjf,CMS:2023gfb} is a multipurpose, nearly hermetic detector, designed to trigger on~\cite{CMS:2020cmk,CMS:2016ngn,CMS:2024aqx} and identify electrons, muons, photons, and (charged and neutral) hadrons~\cite{CMS:2020uim,CMS:2018rym,CMS:2014pgm}. Its central feature is a superconducting solenoid of 6\unit{m} internal diameter, providing a magnetic field of 3.8\unit{T}. Within the solenoid volume are a silicon pixel and strip tracker, a lead tungstate crystal electromagnetic calorimeter (ECAL), and a brass and scintillator hadron calorimeter (HCAL), each composed of a barrel and two endcap sections. The silicon tracker measures charged particles within the pseudorapidity range $\abs{\eta} < 3.0$. During the LHC running period when the data used in this Letter were recorded, the silicon tracker consisted of 1856 silicon pixel and 15\,148 silicon strip detector modules~\cite{Phase1Pixel}. Muons are identified and reconstructed over  $\abs{\eta} < 2.4$ using gas-ionization detectors interleaved with the layers of the steel flux-return yoke outside the solenoid. Forward hadron (HF) calorimeters, made of steel and quartz fibers, extend the $\abs{\eta}$ coverage from 3.0, provided by the barrel and endcap detectors, to 5.2~\cite{CMS:2019hpr}. The HF calorimeters are segmented to form $\Delta\eta{\times}\Delta\phi$ ``towers'' of width $0.175{\times}0.175$, with $\phi$ being the azimuthal angle. Two zero-degree calorimeters (ZDCs), made of quartz fibers and plates embedded in tungsten absorbers, are used to detect neutrons from nonexclusive and UPC EMD events in the region $\abs{\eta} > 8.3$~\cite{Grachov:2006ke}.

Events are selected online using a two-tiered trigger system. The first level, composed of custom hardware processors, uses information from the calorimeters and muon detectors~\cite{CMS:2020cmk}. The second level, known as the high-level trigger~\cite{CMS:2016ngn}, consists of a farm of processors running a version of the full event reconstruction software.

A particle-flow algorithm~\cite{CMS:2017yfk} aims to reconstruct and identify each individual particle in an event, with an optimized combination of information from the various elements of the CMS detector. The energy of photons is obtained from the ECAL measurement. The energy of an electron is determined from a combination of the electron momentum at the primary interaction vertex as determined by the tracker, the energy of the corresponding ECAL cluster, and the energy sum of all bremsstrahlung photons spatially compatible with originating from the electron track. The energy of a muon is obtained from the curvature of the corresponding track. Matching muons to tracks measured in the silicon tracker results in a relative transverse momentum resolution, for muons with \pt up to 100\GeV, of 1\% in the barrel and 3\% in the endcaps. The energy of a charged hadron is determined from a combination of the momentum measured in the tracker and the matching ECAL and HCAL energy deposits, corrected for the response function of the calorimeters to hadronic showers. The energy of neutral hadrons is obtained from the corresponding corrected ECAL and HCAL energies. The reconstructed particles mentioned above are used to build \PGt lepton candidates~\cite{CMS:2018jrd} and to measure the missing \pt of the event~\cite{CMS:2016lmd,CMS:2019ctu}.

\section{Monte Carlo event simulation}
\label{sec:MC}

The $\text{PbPb}\xrightarrow{\gaga}\Pb^{(*)}\tautau\Pb^{(*)}$ signal is generated at leading order (LO) in QED with the \gammaUPC~1.0~\cite{Shao:2022cly} code combined with \MGvATNLO~(v3.5)~\cite{MadGraph5} for the corresponding matrix-elements calculation using the SM \atau value. The \gammaUPC\ charged form-factor photon fluxes are used with initial transverse momentum $\kt$ to account for their small virtualities~\cite{Shao:2024dmk}.
Signal samples are also generated with the \upcgen\ (tag 23-06-22) event generator~\cite{Burmasov:2021phy}, which incorporates variations of the \atau parameter, with values in the range $-0.1$ to $+0.1$ considered. These two Monte Carlo (MC) event generators agree in both the signal cross section and the \PGt lepton kinematic distributions when using the same \atau value and charged form-factor photon flux. Residual differences in the predicted \PGt lepton differential distributions are at the few percent level in the fiducial phase space of the measurement. The final-state radiation (FSR) of the tau leptons (and of their decay electrons and muons), together with their leptonic and hadronic decays, are simulated with \PYTHIA{8}~(v2.1.2)~\cite{Sjostrand:2014zea}, as well as with \TAUOLA~(v1.1.9)~\cite{Chrzaszcz:2016fte} as an alternative model. The exclusive \ggmumuFSR background, of relevance for the \muoneprong and \mue channels, is simulated with \gammaUPC\ at LO accuracy, whereas the \ggee process, a background to the \ethreeprong final state, is simulated using \textsc{SuperChic}~3.03~\cite{Harland-Lang:2018iur}. All generated events are passed through the \GEANTfour~\cite{Agostinelli:2002hh} detector simulation and reconstructed with the same software as used for the collision data.

\section{Event selection, reconstruction, and systematic uncertainties}
\label{sec:selections}

Exclusive photon-fusion ditau events are characterized by a low number of particles produced in an otherwise empty detector.
Therefore, signal events for the \muoneprong, \muthreeprong, and \mue decay channels are selected online by requiring the presence of a single muon with no explicit \pt threshold, at least one pixel detector track, and no activity in at least one of the HF detectors. 
The online selection for the \ethreeprong channel uses a combination of electron triggers with \pt thresholds of 3 or 5\GeV, each requiring at least one empty HF calorimeter. The single-muon trigger efficiency exceeds 90\% over the full $\eta$ range, whereas the corresponding electron efficiency varies from about 20\% at $\pt\approx 3\GeV$ in the endcap region to above 90\% for electrons with $\pt\gtrsim 10\GeV$ in the ECAL barrel.

Offline, the events are required to have exactly the number of muons, electrons, and charged pions expected for each decay mode, with each track passing the ``high-purity'' requirements~\cite{CMS:2014pgm}. Muons must satisfy the ``soft'' quality criteria~\cite{CMS:2018rym} and have $\pt > 3.5\GeV$ for $\abs{\eta} < 1.2$ or $\pt > 1.5\GeV$ for $1.2 < \abs{\eta} < 2.4$, following the acceptance of the muon detector system. For electron reconstruction, the dedicated low-\pt algorithms of the similar $\gaga\to\epem$ measurement are used~\cite{CMS:2024bnt}. The electron minimum \pt threshold is set to 2.5\GeV in the \mue channel (above the typical ECAL noise levels) and to 4\GeV in the \ethreeprong final state (because of a higher online trigger threshold for this channel). In the \muoneprong channel, the charged hadron must have $\pt > 0.3\GeV$. The tracks that form the ``\threeprong'' candidate in the \muthreeprong and \ethreeprong channels must be identified as charged hadrons by the particle-flow algorithm, pass a \pt threshold of 0.5 (0.3)\GeV for the first (second and third) track, and have a common vertex. The kinematics of the visible hadronic $\PGt$ decay in the \muthreeprong and \ethreeprong channels is reconstructed from the vector sum of the three charged hadrons (assumed to be pions); its \pt must exceed 2 and 1\GeV in the \muthreeprong and \ethreeprong channels, respectively, and its invariant mass must be lower than 1.5\GeV in both decay modes.

To further suppress background processes from hadronic PbPb and $\PGg$Pb collisions, events with activity in any of the HF calorimeters, above thresholds that reject 90\% of the detector noise, are discarded. Specifically, events containing at least one HF calorimeter tower with an energy above 5.7--7.0\GeV are rejected in all four final states, where the exact threshold depends on the instantaneous luminosity for \muoneprong and \muthreeprong channels. The small signal efficiency losses induced by the HF selection requirements are corrected using simulation, with residual differences constrained in data control regions, as described in Ref.~\cite{CMS:2024bnt}. Additional exclusivity requirements are applied in the \mue channel by rejecting events that also have any reconstructed charged particle, other than the electron and muon, as well as extra energy deposits in the ECAL and HCAL above noise thresholds. These more stringent requirements follow closely the charged and neutral exclusivity criteria applied in the similar $\gaga\to\epem$ analysis~\cite{CMS:2024bnt}. In the \muoneprong channel, which is more affected by $\PGg$Pb backgrounds than other decay modes, events with activity above noise in either ZDC are vetoed. The requirement of zero neutrons emitted in both directions (0n0n) reduces the \muoneprong signal yield by ${\approx}30\%$, while reducing backgrounds by ${\approx}75\%$. The use of the forward neutron information reduces the \muoneprong data sample to events with fully operational ZDC detectors, lowering the integrated luminosity to 1.61\nbinv for this channel. The measured 0n0n emission probability as a function of dilepton mass in UPCs is well reproduced by the MC simulations~\cite{Harland-Lang:2023ohq,Crepet:2024dvv}. Nevertheless, the signal loss associated with the 0n0n selection is estimated directly from data by measuring the neutron emission probabilities in a control sample of \PbggmmPbPure events, selected under the same criteria, as a function of the dimuon invariant mass. This probability is then used as a weight in the MC simulation of the signal events for the corresponding $\tautau$ invariant masses.

Different photon-induced UPC processes can produce final states with just one reconstructed muon or electron plus a few other charged particles that can mimic the exclusive $\tautau$ signal. Such background sources are suppressed using kinematic selections tailored to each final state. A particularly discriminant quantity is the acoplanarity variable defined as $\Aco^{\Pell\Pell} = 1-\abs{\Delta\phi^{\Pell\Pell}}/\pi$, where $\abs{\Delta\phi^{\Pell\Pell}}$ is the azimuthal angle separation between two leptons. On the one hand, the exclusive $\gaga\to\Pell^+\Pell^-$ processes are characterized by a pair of back-to-back leptons produced with very small azimuthal acoplanarity, $\Aco^{\Pell\Pell} \approx 0$, whereas final states from hadronic and other photon-induced interactions feature larger acoplanarities between the two selected tracks. On the other hand, the visible final states from $\gaga\to\PGtp\PGtm$ decays typically exhibit larger acoplanarities and momentum imbalances, because of escaping neutrinos, than the $\gaga\to\epem,\mumu$ backgrounds. Therefore, the acoplanarity between the muon or electron and the  tau lepton candidates reconstructed in the 1-prong or 3-prong decay modes is required to be lower than 0.1 in the hadronic decay channels, and lower than 0.15 between the two leptons in the \mue mode. In addition, in the \mue channel, the exclusive dimuon background is suppressed by further requiring the acoplanarity between the two tracks to exceed 0.01. In the \muoneprong channel, an additional requirement that the vector sum of the \pt of the two tracks must be larger than 1\GeV suppresses the contamination from $\ggmumu(\PGg)$ processes by more than 91\%. The kinematic selection criteria applied for all decay channels are summarized in Table~\ref{tab:selection}.

\begin{table*}[htpb!]
\centering
\topcaption{Summary of kinematic selection criteria applied in the four $\gaga\to\tautau$ decay channels analyzed. The ``$\PGp$,lead'' and ``$\PGp$,sublead'' labels refer to the leading (highest \pt) and any subleading reconstructed charged hadrons or ``prongs'' (the second- and third-highest \pt particles, assumed to be pions).
\label{tab:selection}}
\renewcommand{\arraystretch}{1.6}
\cmsTable{
\begin{tabular}{cccc}
\hline
{\muoneprong} & {\muthreeprong} & {\mue} & {\ethreeprong} \\
\hline
\multicolumn{3}{c}{\begin{tabular}{c}
$\pt^\PGm > 3.5 \GeV$ for $\abs{\eta^\PGm} < 1.2$ \\
$\pt^\PGm > 1.5 \GeV$ for $1.2 < \abs{\eta^\PGm} < 2.4$
\end{tabular}} & \NA \\
\NA & \NA & $\pt^\Pe > 2.5 \GeV$ & $\pt^\Pe > 4 \GeV$ \\
$\pt^{\PGp} > 0.3 \GeV$ & $\pt^{\PGp,\text{lead}} > 0.5 \GeV$ & \NA & $\pt^{\PGp,\text{lead}} > 0.5 \GeV$ \\
\NA & $\pt^{\PGp,\text{sublead}} > 0.3 \GeV$ & \NA & $\pt^{\PGp,\text{sublead}} > 0.3 \GeV$ \\
\NA & $m_{3\PGp} < 1.5 \GeV$, $\pt^{3\PGp} > 2 \GeV$ & \NA & $m_{3\PGp} < 1.5 \GeV$, $\pt^{3\PGp} > 1 \GeV$ \\
$\Aco^{\PGt_\PGm\PGt_\text{1-prong}} < 0.1$ & $\Aco^{\PGt_\PGm\PGt_\text{3-prong}} < 0.1$  & $0.01 < \Aco^{\PGm\Pe} < 0.15$ & $\Aco^{\PGt_\Pe\PGt_\text{3-prong}} < 0.1$ \\
$\pt^{\PGt_\PGm\PGt_\text{1-prong}} > 1 \GeV$ & \NA & \NA & \NA \\
\hline
\end{tabular}
}
\end{table*}

After all selections, exclusive dimuon production accompanied by FSR, \ggmumuFSR, remains a small source of background in the \muoneprong channel, particularly when one muon fails to reach the muon chambers and is reconstructed as a charged pion. Similarly, a small contribution from \ggee is expected in the \ethreeprong channel when electron bremsstrahlung and photon conversions within the detector material lead to additional reconstructed tracks. The contributions in the signal region for both processes are estimated with the dedicated simulated MC samples discussed in Section~\ref{sec:MC}. To accurately normalize the \ggmumuFSR background, its simulated cross section is scaled to match the yield observed in a control region (CR) defined by events in which the vector sum of the photon momentum and the momenta of the two tracks is close to zero, consistent with exclusive photon-fusion kinematics. The normalization uncertainty from the fit to the CR yield is propagated to the signal region and is found to be negligible, as the residual $\gaga \to \PGm\PGm\PGg$ contribution accounts for less than 10\% of the total background. 

Any residual backgrounds in the \muoneprong, \muthreeprong, and \ethreeprong categories are directly estimated from CRs in data using the ``ABCD'' method~\cite{Behnke:2013pga}. The phase space for these three decay channels is partitioned into four regions defined using two observables that are uncorrelated for background events: the number of reconstructed tracks and the level of HF activity. The track multiplicity is classified as low (1 track in \muoneprong, 3 tracks in \muthreeprong and \ethreeprong) or high (2--3 tracks in \muoneprong, 5--8 tracks in \muthreeprong). The HF activity is defined as low when the maximum HF signal in the event is below the 90\% noise threshold, and high when it lies between the 90 and 99\% thresholds. These thresholds are determined separately for three equal-luminosity data-taking periods, corresponding to 5.7, 6.2, and 7.0\GeV (90\%) and 8.9, 9.7, and 10.6\GeV (99\%). Background contributions in the signal region---defined by low track multiplicity and low HF activity---are then estimated on a bin-by-bin basis using scaled yields from the three CRs.

\begin{figure*}[htbp!]
  \centering
  \includegraphics[width=0.49\textwidth]{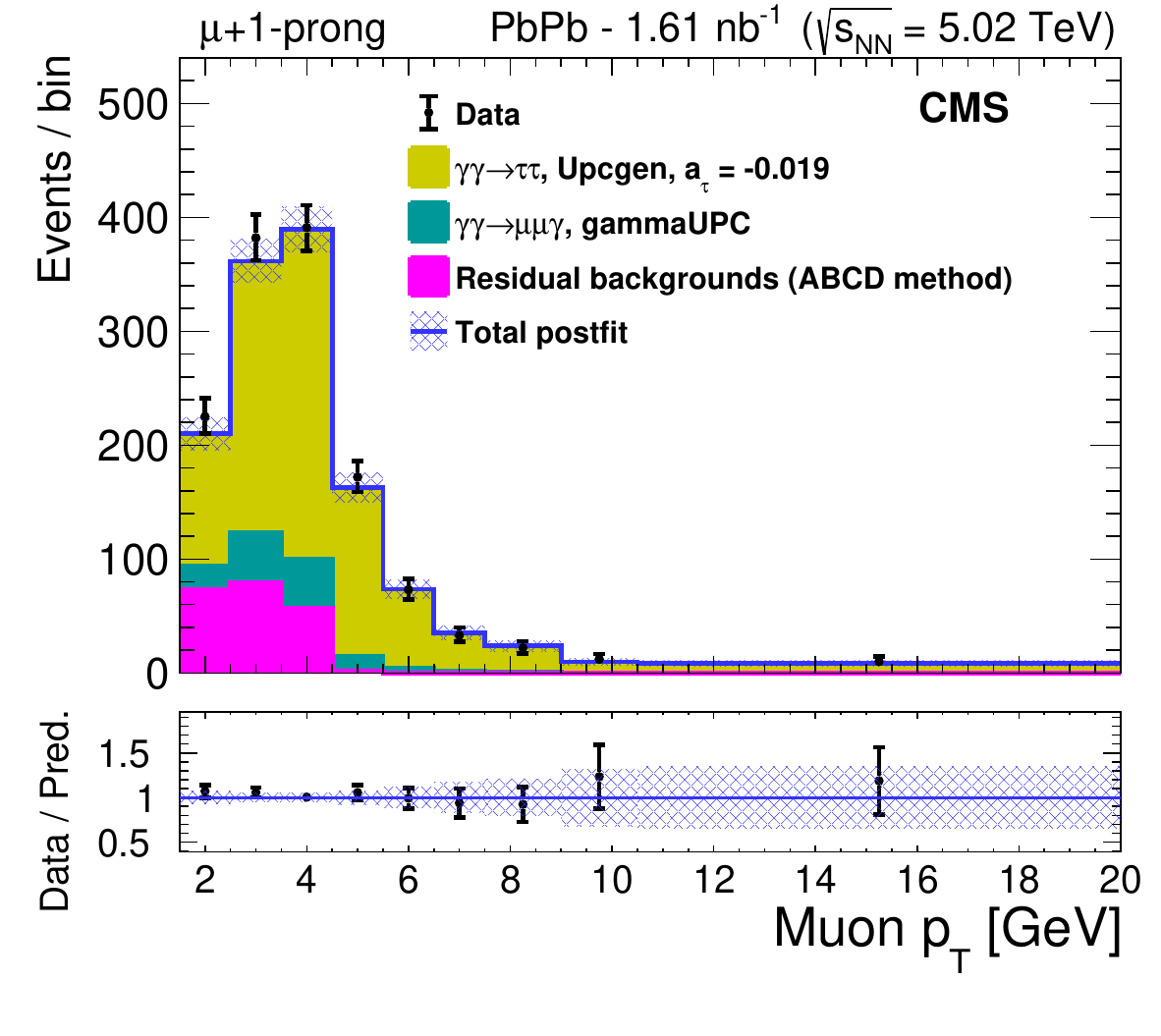}
  \includegraphics[width=0.49\textwidth]{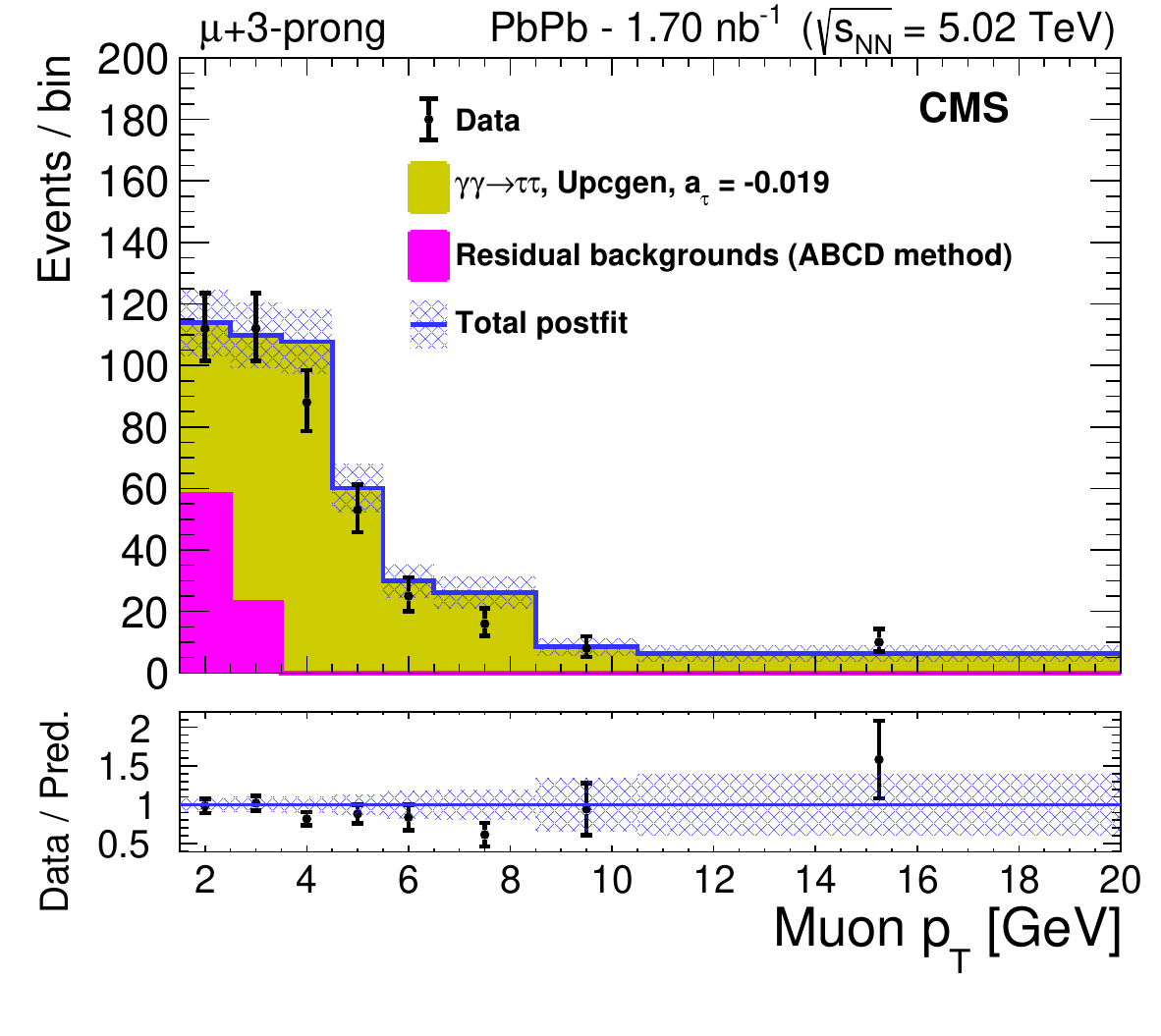}
  \includegraphics[width=0.49\textwidth]{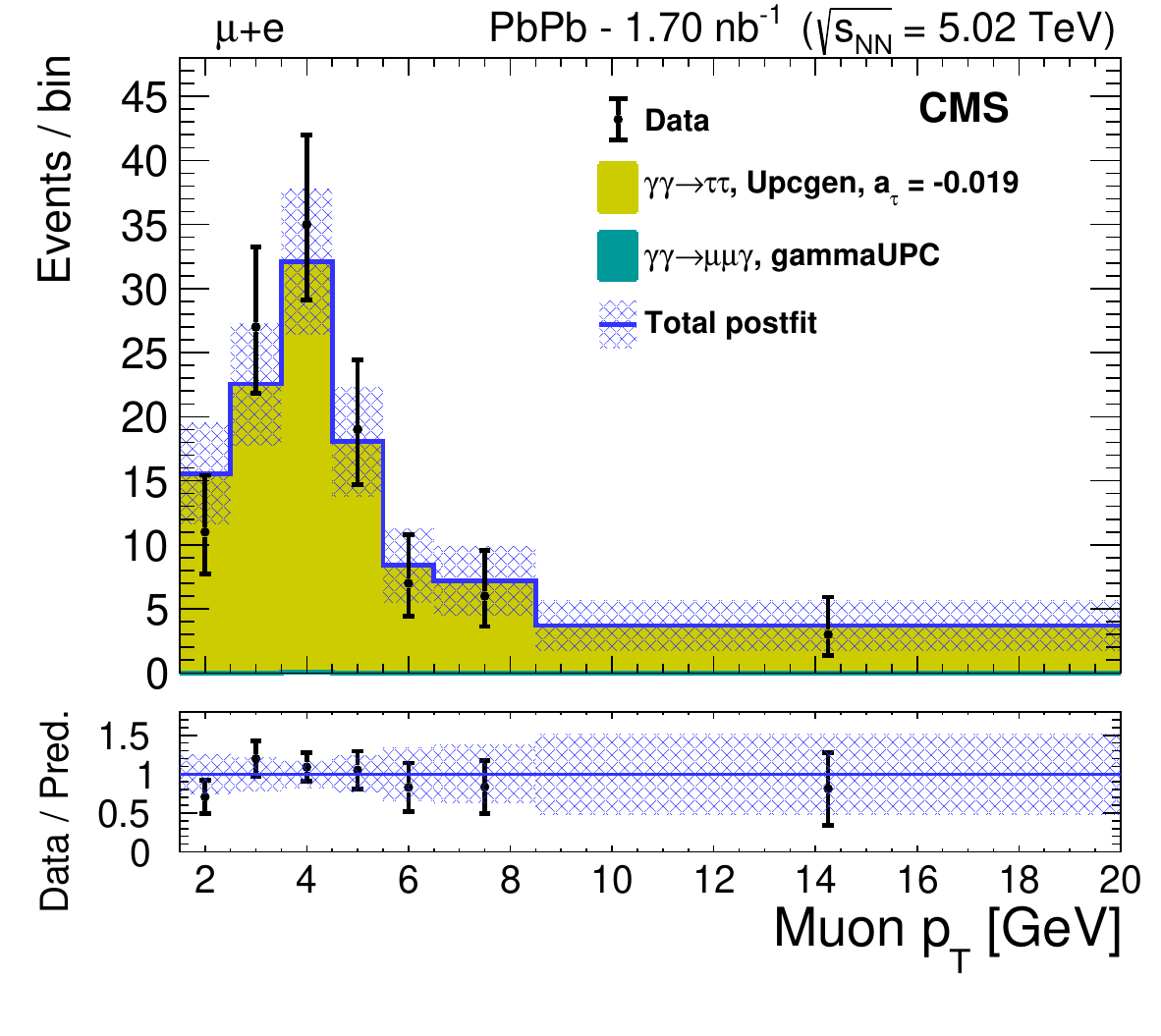}
  \includegraphics[width=0.49\textwidth]{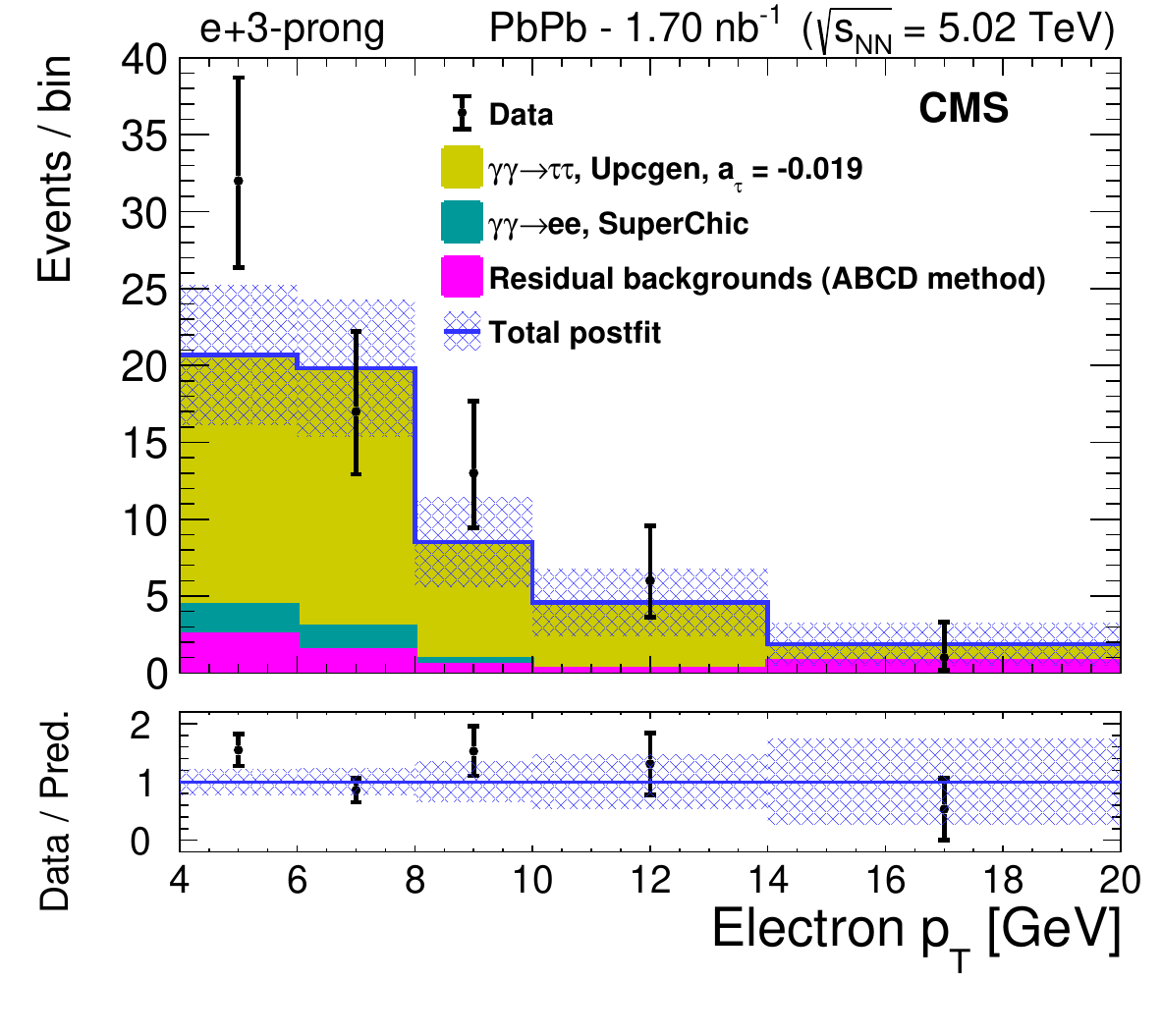}
  \caption{Postfit distributions of the decay charged lepton \pt for the \muoneprong (upper left), \muthreeprong (upper right), \mue (lower left), and \ethreeprong (lower right) channels. The data (black points with error bars indicating statistical uncertainties) are compared with the predicted $\gaga\to\tautau$ signal for the fitted $\atau = -0.019$ value (dark yellow histogram) plus the \ggmumuFSR or \ggee (dark green histograms) and residual (pink histogram) backgrounds. The lower panels show the ratio of the data to the sum of signal plus backgrounds. The dashed blue bands show the total postfit uncertainty in the signal-plus-background.}
  \label{fig:kinematics-pt}
\end{figure*}

The reconstructed \pt distributions of the decay muon or electron in the events passing the four $\gaga\to\tautau$ channel selections are shown in Fig.~\ref{fig:kinematics-pt}. Migrations among reconstructed $\PGtp\PGtm$ decay channels can occur, particularly in the \muoneprong category where only 68\% of the selected events originate from true \muoneprong decays, and 28\% arise from fully leptonic $\PGtp\PGtm$ decays with one lepton misidentified as a charged pion, according to the MC simulations.
In all cases, the measured distributions are well reproduced by the predictions from the simulated $\gaga\to\tautau$ signal (for the best-fit value $\atau= -0.019$, as described in the next section) plus the background contributions, including the exclusive \ggmumuFSR and \ggee processes, as well as the residual background derived from data CRs.

Systematic uncertainties affecting the signal extraction efficiency are derived following the techniques used in previous similar analyses, as described in Refs.~\cite{CMS:2022arf,CMS:2024bnt}. The dominant uncertainties affecting the signal reconstruction, expressed as percentage variations of the cross section measurement and calculated with \atau fixed to $\atau^\text{SM}$ in the fit, arise from the muon trigger and reconstruction efficiencies (3.7\%), the pion tracking efficiency (3.2\%), the HF veto trigger efficiency (2.7\%), differences between the \PYTHIA{8} and \TAUOLA modeling of the $\PGt$ decays and FSR (2.5\%), and the integrated luminosity (1.7\%)~\cite{CMS:2025rzq}. Additional smaller uncertainties arise from the choice of ABCD CRs, limited size of MC event samples, $\PGt$ branching fractions, electron reconstruction efficiency, exclusivity requirements, and the neutron multiplicity categorization (for the \muoneprong channel). Due to cross-feeding among different decay channels and common sources of systematic uncertainties, their impact on the extracted $\tautau$ observables must be assessed in a combined manner. All sources of systematic uncertainties are incorporated into the signal-plus-background model used in the log-likelihood fit performed to extract the $\gaga\to\tautau$ cross section and $\atau$ constraints, as described below.

\section{Fit strategy and results}
\label{sec:results}

\subsection{Likelihood construction}
\label{sec:fit}

Using the CMS \textsc{combine} tool~\cite{CMS:2024onh}, a binned likelihood fit is performed using the four measured decay-lepton \pt distributions shown in Fig.~\ref{fig:kinematics-pt} to simultaneously extract $\atau$ and the $\gaga\to\tautau$ fiducial cross section ($\sigma^\text{fid}$) within the $\pt^\PGt > 1\GeV$ and $\abs{\eta^\PGt} < 3$ phase space. For each bin $i$, the number of observed events $n_i$ is assumed to follow a Poisson probability distribution. The likelihood is defined as
\begin{equation}
L(\atau, \sigma_{\text{fid}}, \boldsymbol{\theta}) 
= \prod_{i} \mathrm{Poiss}\!\left(n_i \mid \nu_i(\atau, \sigma_{\text{fid}}, \boldsymbol{\theta})\right) \prod_{k} \pi_k(\theta_k),
\end{equation}
where $\nu_i$ denotes the predicted yield in bin $i$, and $\boldsymbol{\theta}$ is the set of nuisance parameters describing systematic uncertainties. Each nuisance parameter $\theta_k$ is constrained by a prior term $\pi_k(\theta_k)$. Uncertainties that affect only the overall event yield, such as the luminosity and background normalizations, are implemented with log-normal prior constraints. Shape-related uncertainties, including those from reconstruction and identification efficiencies and from the modeling of the signal acceptance and kinematic distributions, are implemented with Gaussian prior constraints. The impact of these uncertainties in $\nu_i$ is propagated either through alternative ``up'' and ``down'' templates for shape-related effects or through overall yield variations for uncertainties that affect only the normalization. The fit describes the data in all four channels, with almost all data points lying within the postfit uncertainty bands (Fig.~\ref{fig:kinematics-pt}). 

Signal templates are obtained from a smooth bin-by-bin interpolation between 21 MC samples simulated with the \upcgen\ event generator covering the $-0.1 < \atau < 0.1$ range, so as to provide fine granularity in the likelihood scan. The negative log-likelihood is computed over a grid of $\atau$ and $\sigma^\text{fid}$ values, incorporating systematic uncertainties as nuisance parameters. The predicted signal contribution in each bin is determined as
\begin{equation}
\nu_i^{\text{sig}}(\atau, \sigma_{\text{fid}})= \sigma_{\text{fid}}\,\mathcal{L}_{\text{int}}\,A_i(\atau),
\end{equation}
where $\mathcal{L}_{\text{int}}$ is the integrated luminosity and $A_i(\atau)$ denotes the combined acceptance and reconstruction efficiency in bin~$i$, determined by the interpolated signal templates. The two-dimensional (2D) likelihood surface $L(\atau, \sigma_{\text{fid}})$ is evaluated on a dense grid in both parameters. Confidence intervals are derived using the profile likelihood ratio
\begin{equation}
\label{eq:likelihood-ratio}
q(\atau, \sigma_{\text{fid}}) = -2 \ln \frac{L\!\left(\atau, \sigma_{\text{fid}}, \hat{\hat{\boldsymbol{\theta}}}(\atau,\sigma_{\text{fid}})\right)}{L\!\left(\hat{a}_\PGt, \hat{\sigma}_{\text{fid}},\hat{\boldsymbol{\theta}}\right)
},
\end{equation}
where \(\hat{a}_\PGt\), \(\hat{\sigma}_{\text{fid}}\), and \(\hat{\boldsymbol{\theta}}\) denote the unconditional maximum-likelihood estimates, while \(\hat{\hat{\boldsymbol{\theta}}}(a_\PGt,\sigma_{\text{fid}})\) denotes the conditional maximum-likelihood estimate of the nuisance parameters for fixed values of \(a_\PGt\) and \(\sigma_{\text{fid}}\).

\subsection{Simultaneous \texorpdfstring{$\atau$}{a[tau]} and \texorpdfstring{$\sigma^\text{fid}$}{sigma[fid]} extraction}
\label{sec:2Dresults}

\begin{figure}[htb!]
  \centering
\includegraphics[width=0.49\textwidth]{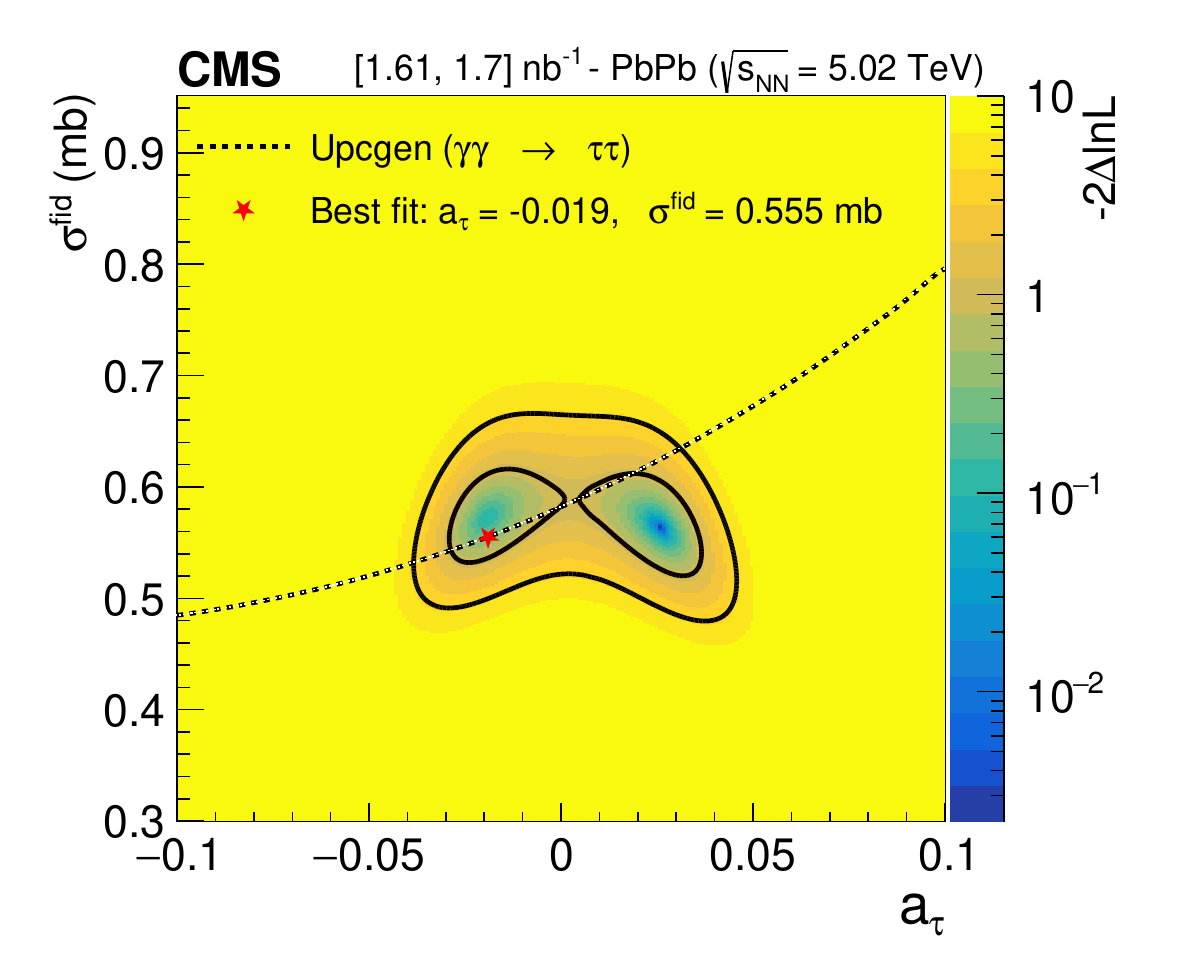}
\includegraphics[width=0.49\textwidth]{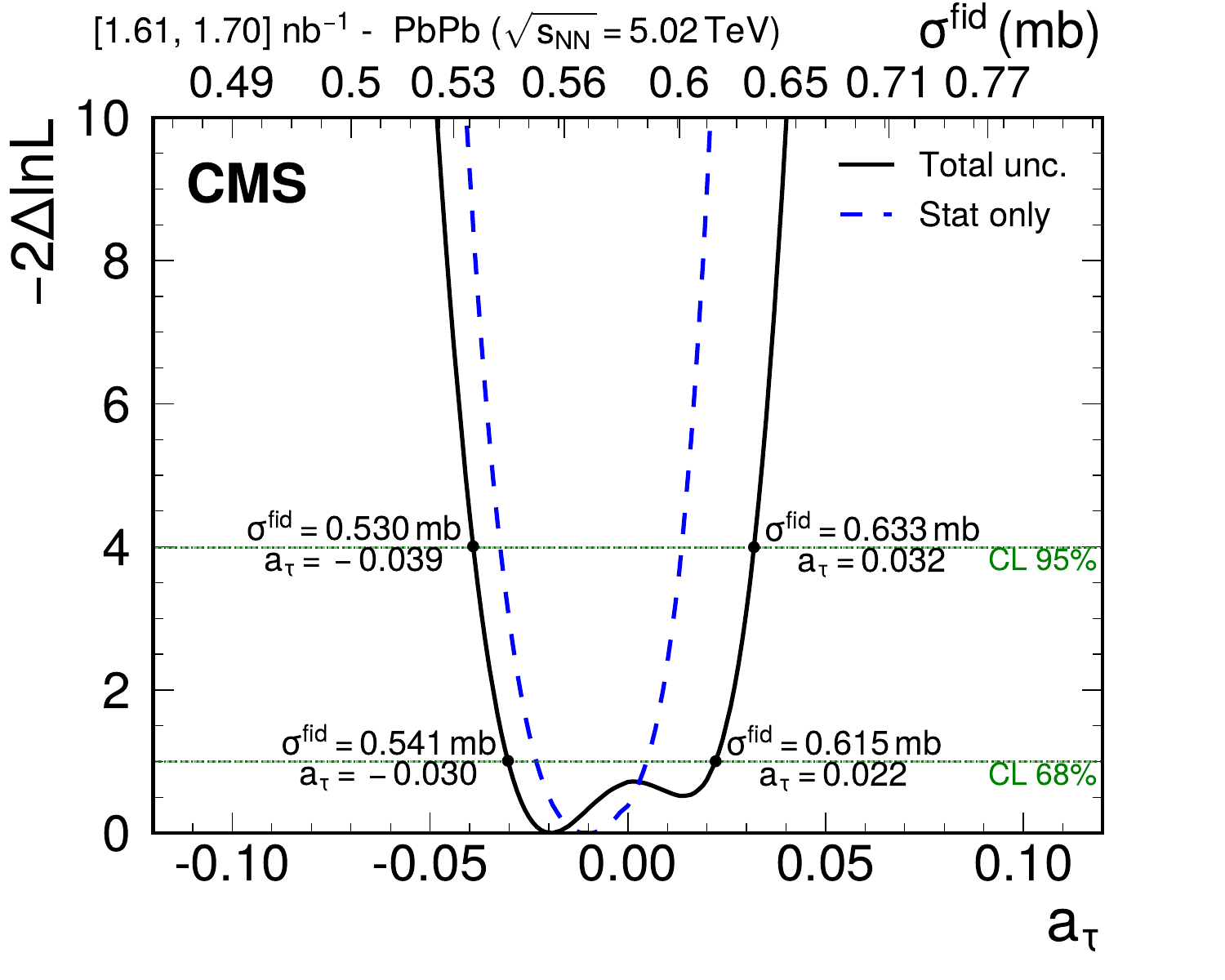}
\caption{Negative log-likelihood results from the combined fit to the decay-lepton \pt distributions of the four \ggtautau channels (Fig.~\ref{fig:kinematics-pt}). \cmsLLeft: Likelihood surface in the $(\atau,\sigma^\text{fid})$ plane with contours at $-2\Delta\ln L = 1$ and 4, \upcgen\ prediction between $\atau$ and $\sigma^\text{fid}$ (black dotted curve, normalized to data), and best fit point along this curve (red star). \cmsRRight: $-2\Delta\ln L$ distributions for $\atau$ (lower $x$ axis) and $\sigma^\text{fid}$ (upper $x$ axis), evaluated along the dotted curve in the left panel, accounting for all uncertainties (black curve) and statistical uncertainties alone (blue dashed curve). The horizontal lines indicate the corresponding 68 and 95\%~\CL intervals.
\label{fig:limits_2D_1D}}
\end{figure}

The negative log-likelihood results from the combined fit to the four \ggtautau decay channels, using their lepton \pt distributions, are shown in Fig.~\ref{fig:limits_2D_1D}. The \cmsLeft panel shows the likelihood-ratio $-2\Delta\ln L$ surface in the $(\atau,\sigma^\text{fid})$ plane, Eq.~(\ref{eq:likelihood-ratio}), with the inner and outer contours at $-2\Delta\ln L = 1$ and 4, respectively, chosen so that their intersections with the dotted curve match the 68 and 95\% confidence level (\CL) intervals of the one-dimensional likelihood. The dotted curve shows the predicted correlation from \upcgen\ between $\atau$ and $\sigma^\text{fid}$, normalized to data, and the red star marks the best fit point along this curve. The \cmsRight panel shows the $-2\Delta\ln L$ distribution evaluated along the dotted curve, with the corresponding 68 and 95\%~\CL intervals. The two nearby minima arise from the  approximate symmetry under $\atau \to -\atau$, which is broken once the fiducial cross section---monotonically increasing with $\atau$---is included.

\subsection{Comparison of \texorpdfstring{$\atau$}{a[tau]} across channels and systems}
\label{sec:atauresults}

\begin{figure*}[htb]
  \centering
  \includegraphics[width=0.8\textwidth]{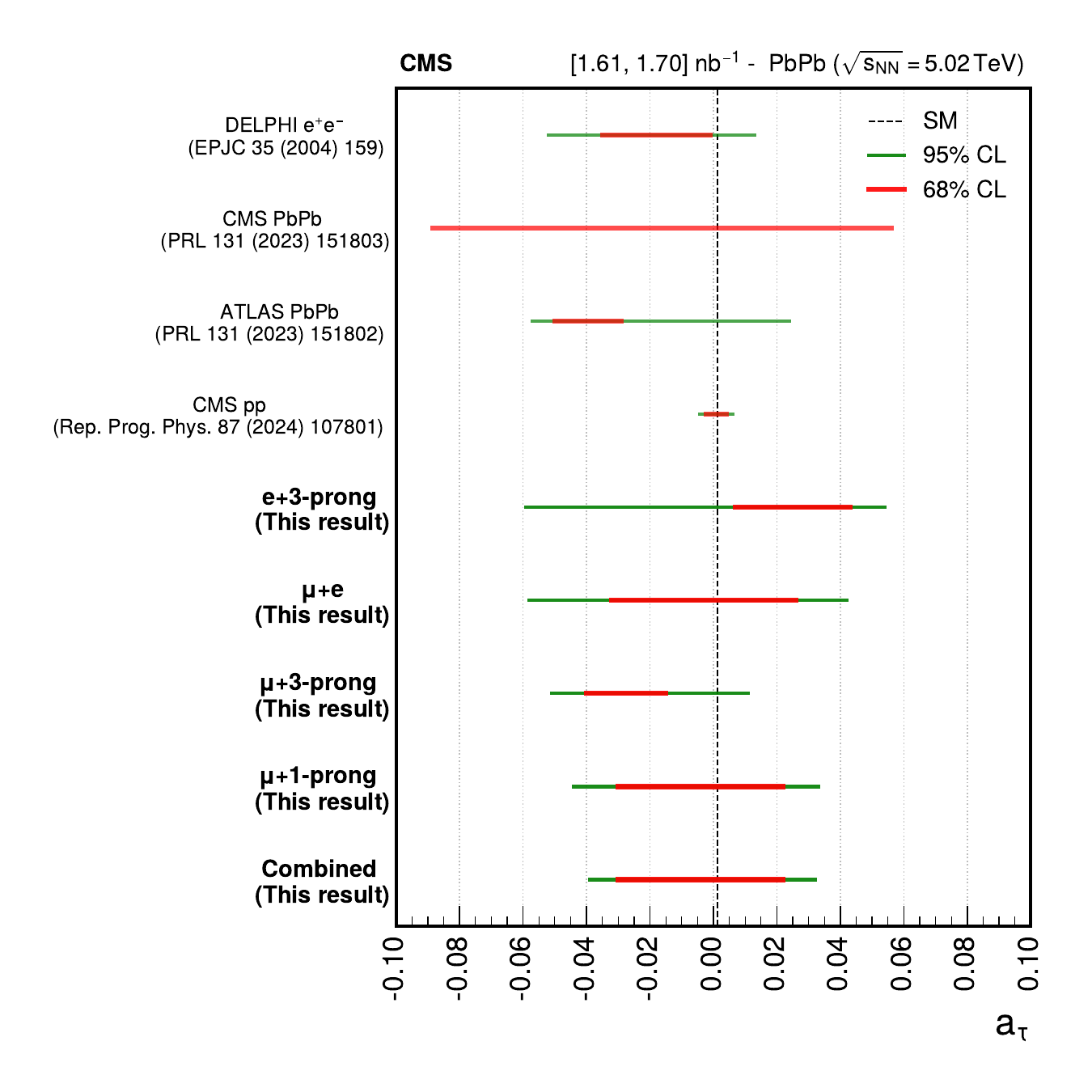}
  \caption{The $\atau$ limits derived in this work for each individual decay channel and their combined result (five lower entries) compared with previous CMS~\cite{CMS:2022arf,CMS:2024qjo}, ATLAS~\cite{ATLAS:2022ryk}, and DELPHI~\cite{DELPHI:2003nah} measurements. The red and green bars indicate their corresponding 68 and 95\%~\CL intervals. The SM value is indicated by the vertical dashed line.}
  \label{fig:comparison}
\end{figure*}

Figure~\ref{fig:comparison} compares the extracted $\atau$ limits for the four individual channels and their combined result (five lower entries) with previous measurements (four upper entries). Among all channels, the \muoneprong mode, with its largest data sample, dominates the combined sensitivity to $\atau$.
Including all four channels improves the 95\%~\CL interval for the combined result, whereas the 68\%~\CL limit is essentially determined by the \muoneprong measurement alone, as it encompasses the symmetric two minima observed in the 2D likelihood (Fig.~\ref{fig:limits_2D_1D}). A similar double-minimum structure is also observed in the ATLAS analysis~\cite{ATLAS:2022ryk}, reflecting the same $\atau \to -\atau$ kinematic degeneracy. In the ATLAS result, the secondary minimum lies outside the 68\%~\CL interval, whereas in the present data it lies within that interval, a difference consistent with statistical fluctuations. The strongest current constraint on $\atau$ is obtained from the photon-fusion production of $\PGt$ pairs measured by CMS in $\Pp\Pp$ collisions at high invariant masses $m_{\tata}>50\GeV$, which have an enhanced sensitivity to $\atau$ variations~\cite{CMS:2024bnt}. The limit derived in the present analysis, \limitatau~at 95\% \CL, is obtained from the $m_{\tata}\gtrsim 5\GeV$ region and achieves a sensitivity comparable to that of the ATLAS~\cite{ATLAS:2022ryk} and DELPHI~\cite{DELPHI:2003nah} results, while improving upon the previous CMS PbPb measurement by more than a factor of three.

\subsection{Measured \texorpdfstring{$\sigma^\text{fid}$}{sigma[fid]} and theoretical interpretation}
\label{sec:sigfidresults}

In addition to the $\atau$ constraints, this study provides a measurement of the $\ggtautau$ fiducial cross section, \resultxsection when fitted simultaneously with $\atau$, and \resultxsectionSM when $\atau$ is fixed to the SM value $\atau^\text{SM} = 0.00117721$. The asymmetric and enlarged uncertainty obtained when $\atau$ is left floating arises from the double-minimum structure of the likelihood, visible in Fig.~\ref{fig:limits_2D_1D}. The systematic component of the uncertainty in $\sigma^\text{fid}$ can be determined from the quadrature difference between the full profile likelihood and that obtained using only statistical uncertainties. Overall, the cross section uncertainty is dominated by systematic sources, with the statistical precision being ${\pm}3.5\%$ when the value of $\atau$ is left free and ${\pm}3.1\%$ when it is fixed in the fit. The measured $\gaga\to\tautau$ cross section has the smallest uncertainty among all similar LHC measurements to date, although the high $\tautau$ mass analysis in $\Pp\Pp$ collisions of Ref.~\cite{CMS:2024qjo} achieves a more precise limit on $\atau$. For the $\atau = \atau^\text{SM}$ case, the measured fiducial cross section agrees with the \upcgen\ and \gammaUPC\ predictions at LO accuracy in QED, $\sigma^\text{fid}_\text{LO} = 568 \pm 3\mub$, as well as with the \gammaUPC\ NLO QED prediction, $\sigma^\text{fid}_\text{NLO} = 570 \pm 3\mub$~\cite{Shao:2024dmk,Shao:2025bma}. The quoted theoretical uncertainties are associated with the modeling of the Pb survival probability and are obtained by varying the nuclear radius and diffusivity of the Pb ion with a Glauber MC model~\cite{Shao:2024dmk,dEnterria:2020dwq}. No extra uncertainty related to the choice of the photon flux is included since the theoretical $\PGg$ spectrum derived from the charged form-factor of the Pb ion, used in the calculations, is known to accurately describe all differential distributions of exclusive dilepton production in UPCs at the LHC~\cite{Shao:2024dmk,CMS:2024bnt,ATLAS:2020epq}. The NLO QED corrections, which account for most of the quantum contributions to $(g_\PGt-2)$ that result in a nonzero value for $\atau$, increase the fiducial cross section by $+0.35\%$. The impact of NLO weak corrections on the $\gaga\to\tautau$ cross section is much smaller~\cite{Shao:2025bma,Dittmaier:2025ikh}.

\section{Summary}
\label{sec:summary}

The production of tau lepton pairs via photon-photon fusion, $\ggtautau$, has been studied in ultraperipheral lead-lead collisions at a center-of-mass energy of $5.02\TeV$ per nucleon pair. The data sample was collected by the CMS experiment in 2018 corresponding to an integrated luminosity of $1.70\nbinv$. Four $\tautau$ decay final states have been studied: \muoneprong, \muthreeprong, \ethreeprong, and \mue, in events where one muon, one electron, and 0, 1, or 3 charged hadrons (``prongs'') are produced exclusively. A simultaneous likelihood fit to the measured transverse momentum (\pt) distributions of the decay leptons in the four different $\PGtp\PGtm$ final states, which incorporates information from both spectral shape and normalization, constrains the anomalous magnetic moment of the $\PGt$ lepton, $\atau$, and provides the fiducial cross section of the process. A 95\% \CL interval $-0.039 < \atau < 0.032$ is set for $\atau$ in a low-mass $\tautau$ phase space complementary to that probed at higher invariant masses in similar photon-fusion processes in proton-proton collisions at the LHC. This new $\atau$ limit reaches a sensitivity comparable to the existing constraints from electron-positron collisions at LEP. A cross section \resultxsection is measured in a fiducial phase space, corresponding to tau lepton transverse momenta $\pt^{\PGt} > 1\GeV$ and pseudorapidities $\abs{\eta^{\PGt}}<3$. This cross section, the most precisely measured at the LHC for this process to date, is in agreement with next-to-leading-order quantum electrodynamics predictions.

\begin{acknowledgments}
We congratulate our colleagues in the CERN accelerator departments for the excellent performance of the LHC and thank the technical and administrative staffs at CERN and at other CMS institutes for their contributions to the success of the CMS effort. In addition, we gratefully acknowledge the computing centers and personnel of the Worldwide LHC Computing Grid and other centers for delivering so effectively the computing infrastructure essential to our analyses. Finally, we acknowledge the enduring support for the construction and operation of the LHC, the CMS detector, and the supporting computing infrastructure provided by the following funding agencies: SC (Armenia), BMFWF and FWF (Austria); FNRS and FWO (Belgium); CNPq, CAPES, FAPERJ, FAPERGS, and FAPESP (Brazil); MES and BNSF (Bulgaria); CERN; CAS, MoST, and NSFC (China); MINCIENCIAS (Colombia); MSES and CSF (Croatia); RIF (Cyprus); SENESCYT (Ecuador); ERC PRG and PSG, TARISTU24-TK10 and MoER TK202 (Estonia); Academy of Finland, MEC, and HIP (Finland); CEA and CNRS/IN2P3 (France); SRNSF (Georgia); BMFTR, DFG, and HGF (Germany); GSRI (Greece); MATE and NKFIH (Hungary); DAE and DST (India); IPM (Iran); SFI (Ireland); INFN (Italy); MSIT and NRF (Republic of Korea); MES (Latvia); LMTLT (Lithuania); MOE and UM (Malaysia); BUAP, CINVESTAV, CONACYT, LNS, SEP, and UASLP-FAI (Mexico); MOS (Montenegro); MBIE (New Zealand); PAEC (Pakistan); MSHE, NSC, and NAWA (Poland); FCT (Portugal); MESTD (Serbia); MICIU/AEI and PCTI (Spain); MOSTR (Sri Lanka); Swiss Funding Agencies (Switzerland); MST (Taipei); MHESI (Thailand); TUBITAK and TENMAK (T\"{u}rkiye); NASU (Ukraine); STFC (United Kingdom); DOE and NSF (USA).

\begin{sloppypar}
\setlength\emergencystretch{\hsize}
\hyphenation{Rachada-pisek} Individuals have received support from the Marie-Curie program and the European Research Council and Horizon 2020 Grant, contract Nos.\ 675440, 724704, 752730, 758316, 765710, 824093, 101115353, 101002207, 101001205, and COST Action CA16108 (European Union); the Leventis Foundation; the Alfred P.\ Sloan Foundation; the Alexander von Humboldt Foundation; the Science Committee, project no. 22rl-037 (Armenia); the Fonds pour la Formation \`a la Recherche dans l'Industrie et dans l'Agriculture (FRIA) and Fonds voor Wetenschappelijk Onderzoek contract No. 1228724N (Belgium); the Beijing Municipal Science \& Technology Commission, No. Z191100007219010, the Fundamental Research Funds for the Central Universities, the Ministry of Science and Technology of China under Grant No. 2023YFA1605804, the Natural Science Foundation of China under Grant No. 12535004, and USTC Research Funds of the Double First-Class Initiative No.\ YD2030002017 (China); the Ministry of Education, Youth and Sports (MEYS) of the Czech Republic; the Shota Rustaveli National Science Foundation (Georgia); the Deutsche Forschungsgemeinschaft (DFG), among others, under Germany's Excellence Strategy -- EXC 2121 ``Quantum Universe" -- 390833306, and under project number 400140256 - GRK2497; the Hellenic Foundation for Research and Innovation (HFRI), Project Number 2288 (Greece); the Hungarian Academy of Sciences, the New National Excellence Program - \'UNKP, the NKFIH research grants K 131991, K 138136, K 143460, K 143477, K 147557, K 146913, K 146914, K 147048, TKP2021-NKTA-64, and 2025-1.1.5-NEMZ\_KI-2025-00004, and MATE KKP and KKPCs Research Excellence and Flagship Research Groups grants (Hungary); the Council of Science and Industrial Research, India; ICSC -- National Research Center for High Performance Computing, Big Data and Quantum Computing, FAIR -- Future Artificial Intelligence Research, and CUP I53D23001070006 (Mission 4 Component 1), funded by the NextGenerationEU program, the Italian Ministry of University and Research (MUR) under Bando PRIN 2022 -- CUP I53C24002390006, PRIN PRIMULA 2022RBYK7T (Italy); the Latvian Council of Science; the Ministry of Science and Higher Education, project no. 2022/WK/14, and the National Science Center, contracts Opus 2021/41/B/ST2/01369, 2021/43/B/ST2/01552, 2023/49/B/ST2/03273, and the NAWA contract BPN/PPO/2021/1/00011 (Poland); the Funda\c{c}\~ao para a Ci\^encia e a Tecnologia (Portugal); the National Priorities Research Program by Qatar National Research Fund; MICIU/AEI/10.13039/501100011033, ERDF/EU, ``European Union NextGenerationEU/PRTR", projects PID2022-142604OB-C21, PID2022-139519OB-C21, PID2023-147706NB-I00, PID2023-148896NB-I00, PID2023-146983NB-I00, PID2023-147115NB-I00, PID2023-148418NB-C41, PID2023-148418NB-C42, PID2023-148418NB-C43, PID2023-148418NB-C44, PID2024-158190NB-C22, RYC2021-033305-I, RYC2024-048719-I, CNS2023-144781, CNS2024-154769 and Plan de Ciencia, Tecnolog{\'i}a e Innovaci{\'o}n de Asturias, Spain; the Chulalongkorn Academic into Its 2nd Century Project Advancement Project, the National Science, Research and Innovation Fund program IND\_FF\_68\_369\_2300\_097, and the Program Management Unit for Human Resources \& Institutional Development, Research and Innovation, grant B39G680009 (Thailand); the Eric \& Wendy Schmidt Fund for Strategic Innovation through the CERN Next Generation Triggers project under grant agreement number SIF-2023-004; the Kavli Foundation; the Nvidia Corporation; the SuperMicro Corporation; the Welch Foundation, contract C-1845; and the Weston Havens Foundation (USA).
\end{sloppypar}
\end{acknowledgments}\section*{Data availability} Release and preservation of data used by the CMS Collaboration as the basis for publications is guided by the  \href{https://doi.org/10.7483/OPENDATA.CMS.1BNU.8V1W}{CMS data preservation, re-use and open access policy}.
\bibliography{auto_generated}
\cleardoublepage \appendix\section{The CMS Collaboration \label{app:collab}}\begin{sloppypar}\hyphenpenalty=5000\widowpenalty=500\clubpenalty=5000\cmsinstitute{Yerevan Physics Institute, Yerevan, Armenia}
{\tolerance=6000
A.~Belyaev\cmsorcid{0000-0003-1692-1173}, A.~Hayrapetyan, A.~Tumasyan\cmsAuthorMark{1}\cmsorcid{0009-0000-0684-6742}
\par}
\cmsinstitute{Institut f\"{u}r Hochenergiephysik, Vienna, Austria}
{\tolerance=6000
W.~Adam\cmsorcid{0000-0001-9099-4341}, L.~Benato\cmsorcid{0000-0001-5135-7489}, T.~Bergauer\cmsorcid{0000-0002-5786-0293}, M.~Dragicevic\cmsorcid{0000-0003-1967-6783}, P.S.~Hussain\cmsorcid{0000-0002-4825-5278}, M.~Jeitler\cmsAuthorMark{2}\cmsorcid{0000-0002-5141-9560}, N.~Krammer\cmsorcid{0000-0002-0548-0985}, A.~Li\cmsorcid{0000-0002-4547-116X}, D.~Liko\cmsorcid{0000-0002-3380-473X}, M.~Matthewman, J.~Schieck\cmsAuthorMark{2}\cmsorcid{0000-0002-1058-8093}, R.~Sch\"{o}fbeck\cmsAuthorMark{2}\cmsorcid{0000-0002-2332-8784}, M.~Shooshtari\cmsorcid{0009-0004-8882-4887}, M.~Sonawane\cmsorcid{0000-0003-0510-7010}, W.~Waltenberger\cmsorcid{0000-0002-6215-7228}, C.E.~Wulz\cmsAuthorMark{2}\cmsorcid{0000-0001-9226-5812}
% ...
\par}
$^{1}$Also at Yerevan State University, Yerevan, Armenia\\
$^{2}$Also at Technische Universit\"{a}t Wien, Vienna, Austria\\
% ...
\end{sloppypar}
%%% END EDITABLE REGION %%%
% skeleton_end
\end{document}